\documentclass[usenatbib]{an}
\usepackage{times}
\usepackage{fancyhdr}
\usepackage{graphicx}
\usepackage{dcolumn}
\usepackage{bm}
\def\be{\beta}

\def\ep{\epsilon}

\def\Ga{\Gamma}

\def\bomega{{\bm{\omega}}}
\def\be{{\mathbf{e}}}

\def\bk{{\mathbf{k}}}

\def\bu{{\mathbf{u}}}

\def\bz{{\mathbf{z}}}
\def\bA{{\mathbf{A}}}
\def\bB{{\mathbf{B}}}
\def\bJ{{\mathbf{J}}}

\def\bna{\mbox{\boldmath $\na$}}
\def\bdot{\mbox{\boldmath $\cdot$}}

\newcommand{\ben}{\begin{equation}}
\newcommand{\een}{\end{equation}}
\newcommand{\bea}{\begin{eqnarray}}
\newcommand{\eea}{\end{eqnarray}}
\newcommand{\ba}{\begin{array}}
\newcommand{\ea}{\end{array}}
\newcommand{\bi}{\begin{itemize}}
\newcommand{\ei}{\end{itemize}}

\def\math{\mathsurround 0pt}
\def\oversim#1#2{\lower.5pt\vbox{\baselineskip0pt \lineskip-.5pt
        \ialign{$\math#1\hfil##\hfil$\crcr#2\crcr{\scriptstyle\sim}\crcr}}}

\def\gap{\mathrel{\mathpalette\oversim {\scriptstyle >}}}

\def\pa{\partial}

\def\half{\frac{1}{2}}
\def\na{\nabla}

\newcommand{\vev}[1]{\langle#1\rangle}

\begin{document}


\title{Scaling laws in decaying helical hydromagnetic turbulence}

\author{Mattias Christensson$^1$, Mark Hindmarsh$^2$, and
Axel Brandenburg$^3$
}%
\institute{%
$^1$ Operational Research, Strategy and Change Division, ECGD,
Harbour Exchange Square, London E14 9GS, UK\\
$^2$ Astronomy Centre, University of Sussex, 
Brighton BN1 9QH, UK\\
$^3$ Nordita, Blegdamsvej 17, DK-2100 Copenhagen, Denmark
}%

\date{Received 29 April 2005; accepted 20 May 2005; published online 1 July 2005}

\abstract{
We study the evolution of growth and decay laws for the magnetic field 
coherence length $\xi$, energy $E_{\rm M}$ and magnetic helicity $H$ 
in freely decaying 3D MHD turbulence. 
We show that with certain assumptions, self-similarity of the magnetic power spectrum alone 
implies that $\xi \sim t^{1/2}$. This in turn implies that magnetic helicity decays as 
$H\sim t^{-2s}$, where $s=(\xi_{\rm diff}/\xi_{H})^2$, in terms of $\xi_{\rm diff}$, the diffusion length scale, and $\xi_{\rm H}$, a length scale defined from the helicity power spectrum.
The relative magnetic helicity 
remains constant, implying that the magnetic energy decays as $E_{\rm M} \sim t^{-1/2-2s}$.  
The parameter 
$s$ is inversely proportional to the magnetic Reynolds number $Re_{\rm M}$,
which is constant in the self-similar regime.
\keywords{MHD -- turbulence --  (cosmology:) early universe}
}

\maketitle


\section{Introduction}

Magnetic fields are ubiquitous in the Universe, being observed in objects 
from planets to galaxy clusters 
(Zeldovich et al.\ 1983, Ruzmaikin et al.\ 1988, Kronberg 1994).
In galaxies and galaxy clusters, the typical 
strength is of order a few $\mu$Gauss, which is thought to be produced by 
dynamo action on a seed field. 
In galaxies the dynamo timescale is roughly a rotation 
period, $10^8$ yr, and a simple calculation 
(Ruzmaikin et al.\ 1988)
based on the age of a typical 
galaxy shows that the seed field must have been about 
\(10^{-20}\) Gauss, or perhaps less in the 
currently favored models with a cosmological term 
(Davis et al.\ 1999).

There is no shortage of ideas for generating this seed field.  The more 
conventional astrophysical explanations are based on a Biermann battery 
operating at the era of reionization (see, e.g., 
Gnedin et al.\ 2000,
and references therein).  There are more speculative ideas based on various 
generation mechanisms in the early Universe 
(Grasso \& Rubinstein 2001),
which have the common feature of producing stochastic, 
homogeneous and isotropic magnetic and velocity fields, 
characterized by their power spectra and initial length scales.
Some of these mechanisms produce stochastic fields with non-zero magnetic helicity 
(Joyce 1997, Cornwall 1997, Vachaspati 2001)
the first of these being maximally helical.
Another common feature of these generation mechanisms in the early Universe 
is that they last for a short time only, typically much less than the time it takes for 
the Universe to double in size at the time of generation, 
after which the magnetic fields 
decay.  As we are discussing mechanisms operating at the era of the electroweak 
phase transition ($10^{-11}\,{\rm s}$) or before, field is generated essentially instantaneously 
compared with any current astrophysical or cosmological timescale.

The subsequent decay of these primordial fields, and also of those in 
star-forming regions, motivates 
the study of  freely decaying magnetohydrodynamic (MHD) turbulence 
(Mac Low et al.\ 1998, Biskamp \& M\"uller 1999, M\"uller \& Biskamp 2000,
Christensson et al.\ 2001).
The decay will not in general just be through linear dissipation, as 
the fields in the early are likely to have high 
 magnetic Reynolds number because of the high conductivity of a 
 fully ionized relativistic plasma (we recall that $Re_{\rm M} = \xi v/\eta$,
 where $\xi$ and $v$ are the typical length scale and velocity of the
 system, and $\eta$ the conductivity).
This has been taken to mean in the cosmological context that the  magnetic field is frozen into 
the plasma, and the scale length of the field increases only with the 
expansion of the Universe. This is in general untrue, because the plasma can 
move, and turbulence 
can transfer energy to different length scales 
(Brandenburg et al.\ 1996).

Thus a study of decaying MHD turbulence is required in order to 
calculate quantities such as the size of the 
seed for the galactic dynamo or the amplitude of the perturbations in the 
temperature of the Cosmic Microwave Background (CMB) radiation 
arising from primordial magnetic field generation 
(Durrer et al.\ 1998, 2003, Caprini \& Durrer 2001, Caprini et al.\ 2003).
Our numerical results 
(Christensson et al.\ 2001)
uncovered decay laws for magnetic fields 
which were not those one would expect from purely linear dissipative processes. 
In particular we saw that helical fields decayed 
more slowly than non-helical, due to the fact that magnetic helicity is an invariant 
of ideal MHD.
Helicity is known to be important in dynamo theory
(Pouquet et al.\ 1976, Meneguzzi et al.\ 1981, Brandenburg 2001),
and we shall also be able to confirm its 
importance in decaying turbulence.
Our interest here is to try and understand the 
results, and to compare them with those found by 
Biskamp \& M\"uller (1999) and M\"uller \& Biskamp (2000).
In doing so we have developed a new framework for understanding 
scaling in decaying 3D MHD 
turbulence, in the case where the fields are close to being maximally helical, 
as in the mechanisms proposed 
by Joyce \& Shaposhnikov (1997) and Vachaspati (2001).
It should be noted that none of the estimates 
by Durrer et al.\ (1998, 2003), Caprini \& Durrer (2001), and Caprini et al.\ (2003)
take into account the decay laws we find, and so our results are of direct importance for  
cosmology.

The decay of the magnetic field in the turbulent case is often presumed 
to result from an inverse cascade
(Pouquet et al.\ 1976, Meneguzzi et al.\ 1981, Brandenburg 2001),
in which power is transferred 
locally in $k$-space from small to large scales.  
However, while it is certainly true that power is transferred from small to large scales, 
as can been seen from the energy power spectra plotted in 
Christensson et al.\ (2001),
it was not established that the power is transferred locally, despite the 
appearance of the term ``inverse cascade'' in the title of our paper,  and so it may 
be strictly incorrect to call the process a cascade.   
In fact, a true inverse cascade seems rather 
unlikely, as large scale power appears almost immediately even from initial conditions 
which are highly localized in $k$-space.
For our analysis in this work 
it will not matter whether or not there is a cascade, and we will not discuss 
the matter further.
However, we emphasize that the decay of the energy is not simple linear dissipation: 
there is definitely interesting non-linear physics, as 
shown by the appreciable Reynolds numbers (100-600) and the processing of the 
initial power spectra.

Various scaling 
arguments have been put forward to obtain the growth law 
for the length scale of the magnetic field 
and the decay law for the energy. 
For ideal MHD (infinite conductivity), Olesen (1997) and later Son (1999), 
Field \& Carroll (2000), 
and Shiromizu (1998) 
argued
\(
\xi(t) \sim t^{2/(n+5)} 
\)
where $t$ is conformal time, and $n$ is the index of the initial magnetic
power spectrum.  
Supporting evidence for this scaling law was also found in two-dimensional 
MHD simulations 
(Galtier et al.\ 1997).
Shiromizu's results were based on a renormalization group 
argument, which has been revisited by Berera \& Hochberg (2001), 
who do not find evidence for an inverse cascade
in the steady state.
The effect of having significant helicity 
was supposed to modify this scaling law to
\(\xi(t) \sim t^{2/3}\), 
\( E_{\rm M} \sim t^{-2/3}\)
(Biskamp 1993, Son 1999, Field \& Carroll 2000).
Early numerical experiments with a shell model of the full MHD equations 
(Brandenburg et al.\ 1996) 
suggested \(\xi \sim t^{0.25}\), and 
gave supporting evidence to the inverse cascade. 
Full MHD simulations by Biskamp \& M\"uller (1999) and M\"uller \& Biskamp (2000)
showed, in the helical case,  
an energy decay law 
\(
E_{\rm M} \sim t^{-1/2},
\)
supported by a phenomenological model, which we will discuss at the end of this work.
Decaying non-helical turbulence was studied semi-analytically in a shell 
model by Basu (2000) 
who found an energy decay law of $E_{\rm M} \sim t^{-1.2}$. 
The decay of helical fields was also studied semi-analytically
by Sigl (2002), 
giving a growth law for the length scale between 
$t^{1/2}$ and $t^{2/3}$. Unfortunately, direct comparison with our results is 
otherwise difficult because the correlation functions are expressed in real space.
The importance of magnetic helicity in slowing down the decay has been
recognized earlier in studies of a low-order model of three-dimensional
hydromagnetic flows (Stribling \& Matthaeus 1991), 
where it was also found that
a finite initial cross helicity (not studied in the present paper)
can slow down the decay.
However, power law exponents of the decay have not been determined for
this low-order model.

In an earlier paper (Christensson et al.\ 2001) 
we performed 3D simulations both with and without magnetic helicity, 
starting from homogeneous and isotropic random initial 
conditions, with power spectra suggested by cosmological applications. 
We found 
that the coherence 
scale of the field grows approximately as $t^{1/2}$, 
with significant transfer of power to small scales in the helical case, 
which we ascribed to an inverse cascade.  
The magnetic power spectrum was self-similar with an approximately $k^{-2.5}$ behavior at 
high $k$.
We found decay laws for the magnetic and kinetic energies of 
$t^{-0.7}$ and $t^{-1.1}$ in the helical case, and $t^{-1.1}$ for both in 
the non-helical case.  These are close to, but not identical to those
found by Biskamp \& M\"uller (1999), 
and we suggested that their relatively large initial length scale, 
25\% of the simulation box size, might account  
for the difference.
It should be emphasized that we are interested in the decay only while the scale length of 
the flows is less than the simulation volume, as we want results relevant to the early Universe 
where there are no boundaries.

In this paper we present a new theoretical understanding of our numerical 
results for the power law behavior of the length scale and the energies 
in the helical case.
 We show that the key to understanding the power laws is the self-similarity of 
 the magnetic field, coupled with the near-invariance of the helicity, 
 and that the crucial parameter controlling the rate of decay 
 of the magnetic energy and the helicity is 
 the magnetic Reynolds number.

Our theoretical model has analogies with decaying fluid turbulence in 
2 dimensions, where 
there is also an ideal invariant, the kinetic energy, which plays 
a similar role to the helicity in 3D. 
Decaying turbulence in 2D (Ting et al.\ 1986, Chasnov 1997) 
exhibits self-similarity, and power-law decays in the kinetic energy and 
enstrophy 
(mean squared vorticity) 
are observed in numerical simulations at high Reynolds number (Chasnov 1997). 
We reserve detailed discussion for Section~\ref{s:TwoDTurb}.

\section{MHD equations}

The matter and radiation in the early Universe is modeled as an 
isothermal compressible gas with a 
 magnetic field, which is governed by the momentum equation, the continuity
 equation, and the induction equation, written here in the form 
(Waleffe 1993)
 \bea
    \frac{\pa\bu}{\pa t} =
    -\bu\cdot\bna\bu&-&
    c_{s}^{2}\bna \ln\rho +
    \frac{\bJ\times\bB}{\rho} \nonumber\\
    &+&\frac{\mu}{\rho}\left(\na^{2}\bu +
     \frac{1}{3}\bna\bna\cdot\bu\right),\label{MHD_v} \\
    \frac{\pa \ln\rho}{\pa t} &=& 
    -\bu\cdot\bna \ln\rho - 
    \bna\cdot\bu,\\
    \frac{\pa\bA}{\pa t} &=& 
    \bu\times\bB + 
    \eta\na^{2}\bA,\label{MHD_A}
\eea
 where $\bB=\nabla\times \bA$ is the magnetic field in terms of the magnetic
 vector potential $\bA$, $\bu$ is the velocity, $\bJ$ is the current density, 
 $\rho$ is the density, $\mu$ is the dynamical viscosity, and $\eta$ is the 
 magnetic diffusivity.
 In an expanding Universe the equations are identical when expressed 
in terms of conformally rescaled fields $\bB$, $\bu$ and dissipation parameters 
$\nu$, and $\eta$.  We work in the gauge $A^0 = -\eta \bna\bdot\bA$ 
(Subramanian \& Barrow1998, Christensson et al.\ 2001)
under the assumption that $\eta$ is uniform.

In Fourier space
it is useful 
 to represent the vector potential in terms of its projection onto an orthogonal
 basis formed by $\hat{\be}_{+}$, $\hat{\be}_{-}$ and $\hat{\bk}$, or
 \ben
 \bA_\bk = A^+_\bk\hat{\be}_{+}+A^-_\bk\hat{\be}_{-}+A^0_\bk\hat{\bk}.
 \een
 The two basis vectors $\hat{\be}_{+}$ and $\hat{\be}_{-}$ can be chosen
 to be the unit vectors for circular polarization, right-handed and left-handed 
 respectively. That is
 $\hat{\be}_{\pm} = \hat{\be}_{1} \pm i\hat{\be}_{2}$
 where $\hat{\be}_{1}$ and $\hat{\be}_{2}$ are unit vectors orthogonal to 
 each other and to $\bk$. They are given by 
 $\hat{\be}_{1} = \bk\times\hat{\bz}/|\bk\times\hat{\bz}|$ and
 $\hat{\be}_{2} = \bk\times(\bk\times\hat{\bz})/|\bk\times(\bk\times\hat{\bz})|$
 respectively. $\hat{\bz}$ is a reference direction.
 
 Note that since  
 \ben
   i\hat{\bk}\times\hat{\be}_{s} = sk\hat{\be}_{s}
 \een 
 where $s = \pm 1$, this corresponds to an expansion of the magnetic
 vector potential into helical modes.
 
 Using these basis vectors it is easily seen that the shell-averaged magnetic energy 
 spectrum is
 \ben
 E_{\rm M}(k) = 2\pi k^{2}\langle|\bB_{\bk}|^{2}\rangle
 \een
 where the amplitude of the magnetic field is given by
 \ben
   | \bB_{\bk}|^{2} = 
   (| A_{\bk}^{+}|^{2} + 
   | A_{\bk}^{-}|^{2}) |\bk|^{2}
 \een 
 and the expression for the shell-averaged magnetic helicity spectrum $H(k)$ is
 \ben
 H(k) = 4\pi k^{2}\langle\bA_{\bk}^{*}\cdot\bB_{\bk}\rangle
 \een
 where
 \ben   
   \bA_{\bk}^{*}\cdot\bB_{\bk} = 
   (| A_{\bk}^{+}|^{2} - 
   | A_{\bk}^{-}|^{2}) |\bk|.
 \een
 The function
 $H(k)$ is bounded 
 in magnitude by the inequality 
 \ben
   | H(k)|\leq 2k^{-1} E_{\rm M}(k).
   \label{hel_bound_2}
 \een 
 A field which saturates the above inequality is maximally helical.  

\section{3D MHD simulations of decaying turbulence}

We solve Eqs.\  (\ref{MHD_v})--(\ref{MHD_A}) numerically (Brandenburg 2001) 
 using a variable third order Runge-Kutta
 timestep and sixth order explicit centered space derivatives. 
 All runs are performed on a $120^3$ grid, using periodic boundary conditions, 
 as is appropriate when modeling an infinite volume system.
The average density
 $\langle \rho\rangle = \rho_{0}$
 (where the brackets denote a volume average) 
  is conserved.
 
We use natural units where the speed of light is
$c=1$, and fix the unit of length by setting 
$k_{1}=1$, where $k_{1}$
is the smallest wave number in the simulation box.  Hence the 
box has size $2\pi$. 
The scale factor is fixed by setting 
$\rho_0=1$, and $\bB$ is measured
in units of $\sqrt{\mu_{0}\rho_{0}} c$, where $\mu_{0}$ is the magnetic
permeability. 
We define the mean kinematic viscosity as $\nu \equiv \mu/\rho_{0}$.
The sound speed $c_{s}= 1/\sqrt{3}$, as appropriate for a relativistic fluid.

The equations are not quite those for a 
relativistic gas in the early universe (Brandenburg et al.\ 1996). 
However, we have 
checked that our results change little when using the true relativistic 
equations in the low velocity limit (Christensson et al.\ 2001). 
 
We take 
$\bu$ and $\bB$ to be homogeneous and isotropic Gaussian random 
fields drawn from a power-law distribution with a high wavenumber cut-off. 
The mechanisms for the production of magnetic fields, such as the helical 
production mechanism of 
Joyce \& Shaposhnikov (1997),
are all stochastic, homogeneous 
and isotropic, and have an associated length scale $k_c^{-1}$.  
Hence the power spectra, 
 $P_{\rm M}(k)\equiv \langle\bB^{*}_\bk\cdot\bB_\bk\rangle$, and
 $P_{\rm V}(k)\equiv \langle\bu^{*}_\bk\cdot\bu_\bk\rangle$
can be initially modeled as 
\bea
P_{\rm M}(k)&=&A_M k^{n_{\rm M}} e^{ - (k/k_c)^4},\\
P_{\rm V}(k)&=&A_V k^{n_{\rm V}} e^{ - (k/k_c)^4}.
\eea
Causality demands that $n_{\rm M}\ge 2$ and 
$n_{\rm V}\ge 0$ 
(Durrer et al.\ 1998, 2003, Caprini \& Durrer 2001, Caprini et al.\ 2003).
Note that in the plots it is the 
shell-integrated energy spectra, 
$E_{\rm M,V}(k)=4\pi k^2\times{\half}P_{\rm M,V}(k)$,
which are shown.

The amplitudes $A_{\bk}^{\pm}$ can be chosen independently, provided
$A^{*\pm}_{-\bk} = A^{\pm}_{\bk}$, which is just the condition that
the vector potential be real.
 Therefore it is possible to adjust the amplitudes $| A_{\bk}^{+}|$ and 
 $| A_{\bk}^{-}|$ freely and in so doing obtaining a magnetic field with 
 arbitrary magnetic helicity. With our 
 method we are able to put statistically random but maximally helical fields 
 in our initial conditions.  In our runs with initial helicity we take 
 $H = H_{\rm max}$.
We took $n_{\rm M,V}$  to have the lowest values
consistent with causality, and chose $k_c = 30$.

The initial magnetic energy was taken 
equal to the kinetic energy, and  had the value $5\times10^{-3}$ in all runs. 
The initial density was uniform and equal to 1.
The values for $\nu$ and $\eta$ for the runs presented here are 
summarized in Table \ref{T1}.

\begin{table}
\centering
\begin{tabular}{cccccccc}       \hline
 Run  & A             & B             & C             & D             & E             & F             & G  \\  \hline
$10^{4}\nu$     & $1.0$ & $0.7$ & $0.7$ & $0.7$ & $0.7$ & $0.7$ & $0.7$  \\
$10^{4}\eta$    & $1.0$ & $0.7$ & $0.6$ & $0.5$ & $0.4$ & $0.3$ & $0.2$   \\  
$vt/\xi_{\rm H}$   & $0.49$  & $0.59$  & $0.62$ & $0.64$ & $0.66$ & $0.67$ & $0.69$   \\  
$10^3Re^{-1}_{\rm M}$  & $9.2$ & $5.4$ & $4.5$ & $3.7$ & $3.0$ & $2.4$ & $1.6$       \\
\hline
\end{tabular}
\caption{\label{T1} Kinematic viscosity $\nu$, magnetic diffusivity $\eta$,
$vt/\xi_{\rm H}$ and the inverse of the
magnetic Reynolds number $Re_{\rm M} = \xi_{\rm H}v/\eta$ for our runs.
$\xi_{\rm H}$ is the helicity length scale defined in Eq.\ (\ref{length_H}) and $v$ is the 
RMS velocity, evaluated at the end of the runs ($t \simeq 80$).}
\end{table}

These values were chosen so as to maximize the 
Reynolds numbers while maintaining numerical stability and resolving the 
dissipation scales $2\pi\sqrt{\nu t}$ and $2\pi\sqrt{\eta t}$. The magnetic dissipation scale
is resolved for times $t  > 1/N^2\eta$, where $N = 120$ is the lattice size, with a similar formula 
for the viscous damping scale.  We start taking data at  $t \simeq 40$, when  
$\xi_{\rm diff} \simeq 2\pi \sqrt{8}$ for the smallest of our resistivity values, which corresponds 
to $k \simeq 30$,  or approximately 3 lattice points. Examining 
Fig.\ 1 of Christensson et al.\ (2001), 
we can see the magnetic energy spectrum turning down 
for $k \gap 30$ at $t=46.3$, showing that the system is exhibiting Ohmic dissipation as required.
Under-resolved runs show ringing in real space and the velocity and the magnetic field explode 
fairly quickly, and so are in practice easy to discard.


\section{Helicity conservation and magnetic energy decay}
\label{helicity_conservation}
In this section we shall take a closer look at the observed scaling laws
for helical magnetic fields and see if we can understand them on theoretical 
grounds.
Magnetic helicity is not exactly conserved if $\eta\neq0$, as it straightforward to show that 
\ben
 {\dot H} =
  -2\eta \int \frac{d^{3}\,k}{(2\pi)^{3}}\,
  k^{3}\left(|A_{\bk}^{+}|^{2} - |A_{\bk}^{-}|^{2}\right)
.\een
Hence, providing the power spectrum of the gauge field decays faster than $k^{-6}$
for $k\to\infty$,
implying that the magnetic energy spectrum $E_{\rm M} (k)$
decays faster than $k^{-2}$, we can define a helicity scale
$\xi_{\rm H}$ such that
\ben
  {\dot H} = -8\pi^2\eta\,H/{\xi_{\rm H}^{2}}
  \label{length_H}
.\een
If we assume that the evolution of $\xi_{\rm H}$ is described by a power law
$\xi_{\rm H} \sim t^{r}$ it is clear that the solutions to Eq.\ (\ref{length_H})
are qualitatively different depending on the exponent $r$.  
If and only if $r\,=\,1/2$ does the magnetic helicity show a power law decay
\ben
H\sim t^{-2s}
\een
where 
\ben
s = (\xi_{\rm diff}/\xi_H)^2,
\een 
in terms of the diffusion scale $\xi_{\rm diff} = 2\pi\sqrt{\eta t}$.
Additional length scales we consider are the integral scale
$\xi_{\rm I} = 2\pi \int dk k^{-1}E_{\rm M}(k) / \int dk E_{\rm M}(k)$, 
the relative helicity scale 
$\xi_{\rm R} = \pi |H| / E_{\rm M}$ and the magnetic Taylor microscale 
$\xi_{\rm T} = 2\pi B_{\rm rms} / J_{\rm rms}$, where $B_{\rm rms}$ and 
$J_{\rm rms}$ are the RMS magnetic field and current density respectively. 
It is plausible that all these scales are proportionally related and 
 Fig.~\ref{length_scales} shows that this is indeed the case.
A theoretical reason for this behavior is given below.
\begin{figure}[t!]
\centering{
\resizebox{8.5cm}{!}{\includegraphics{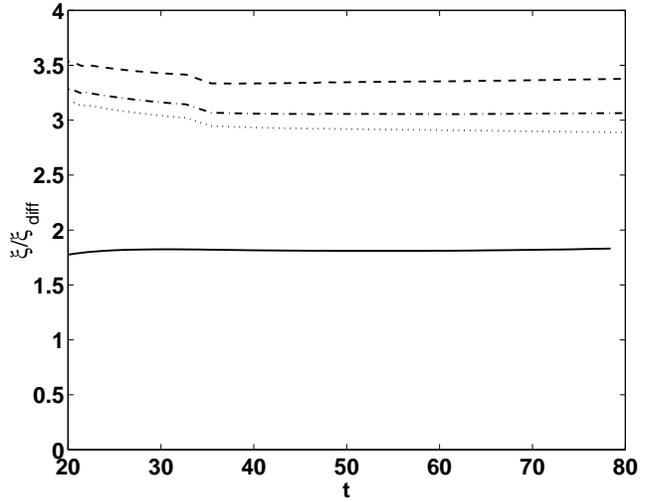}}
}
\caption{Time evolution of the ratio of length scale to the diffusion scale for run D.
The notation is $\xi_{\rm I}/\xi_{\rm diff}$ (dashed),
$\xi_{\rm H}/\xi_{\rm diff}$ (dot-dashed),
$\xi_{\rm R}/\xi_{\rm diff}$ (dotted) and
$\xi_{\rm T}/\xi_{\rm diff}$ (continuous).  The length scales are defined  
in Eq.\  (\ref{length_H}) and the subsequent  
text.
}
\label{length_scales}
\end{figure}

The magnetic helicity is bounded
in magnitude by the inequality Eq. (\ref{hel_bound_2}).
An equivalent way to express this is by
\ben
  H_{\rm REL} \equiv \frac{\pi |H_{\rm M}|}{\xi_{\rm I} E_{\rm M}} \leq 1
  \label{energy_helicity_inequality_2}
,\een 
where $H_{\rm REL}$ is the relative magnetic helicity. 
So if the above bound remains approximately saturated, 
i.e. $H_{\rm REL} \sim 1$, and the helical length scale goes as  
$\xi_{\rm H} \sim t^{1/2}$, the decay law for
the magnetic energy is
\ben
  E_{\rm M} \sim t^{-1/2 - 2s}
  \label{energy_dec}
.\een
In any case, the energy cannot decay faster than this.
Given   $H_{\rm REL}=\xi_{\rm R}/\xi_{\rm I}$, it is seen from Fig.~\ref{length_scales} 
that $H_{\rm REL}$ is indeed of order unity and does not decay markedly with time.

\begin{figure}[t!]
\centering{
\resizebox{8.5cm}{!}{\includegraphics{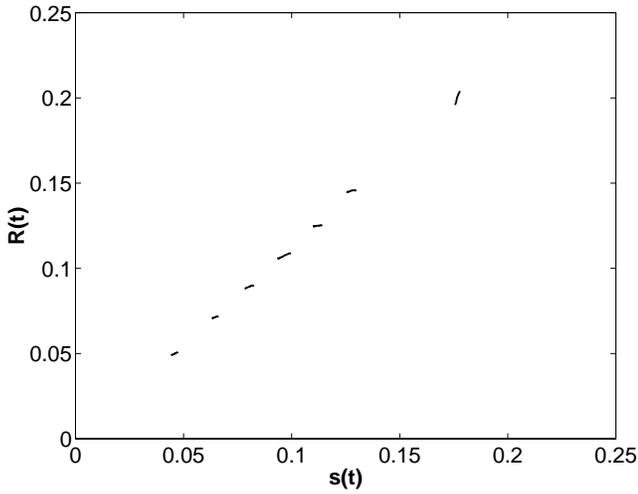}}
}
\caption{The quantity $R(t)$, defined by Eq.\ (\ref{RR}) versus 
the quantity $s(t) = (\xi_{\rm diff}/\xi_H)^2$
for runs listed in Table \ref{T1}. The length scales $\xi_H$ and $\xi_{\rm diff}$ are defined  
in Eq.\  (\ref{length_H}) and the subsequent  
text. The time span is approximately from $t\simeq 40$ to $\simeq 80$,
i.e.\ the last half of the simulations. 
}
\label{s_dot}
\end{figure}

To characterize the decay laws  we define the exponents
\ben
  Q(t) = -t{\dot E}_{\rm M}/E_{\rm M}, \qquad R(t) = -t\dot{H}/2{H}
  \label{RR}
.\een
In Fig.~\ref{s_dot} we have plotted $R(t)$ versus the quantity
$s(t) = (\xi_{\rm diff}/\xi_H)^2$ for several runs with different initial 
conditions. The time span is $t\sim 40$ to $\sim 80$,
i.e.\ the last half of the simulations.
This figure tells us several things. Firstly, it shows us that the value 
of $R$ is approximately independent of time which confirms the power law decay
of $H$. Secondly, 
Fig.~\ref{s_dot} indicates that the quantity $s$ is also approximately 
independent of time, hence reinforcing the relation   
$\xi_H \sim t^{1/2}$. Thirdly, it seems that  $R$ and $s$ are almost
equal, and not just proportional. 

Taking the relative helicity to be constant, 
it follows from the power-law behavior of $H$ that 
the energy decay law is indeed $  E_{\rm M} \sim t^{-1/2 - 2s}.$
In the limit of exact conservation of magnetic helicity, 
$s\rightarrow 0$,  the magnetic energy must decay as $E_{\rm M} \sim t^{-1/2}$.

Regarding the physical significance of the parameter $s$ we note that if
$\xi_{\rm H} \simeq v t$, where $v$ is the RMS velocity,  (i.e.\  if the eddy turn-over time is $t$) 
then $s \simeq (2\pi)^{2} / Re_{\rm M}$, 
where $Re_{\rm M}$ is the magnetic Reynolds number evaluated using the helicity scale 
$\xi_{\rm H}$. 
We have measured $f=vt/\xi_{\rm H}$ and $Re^{-1}_{\rm M}$ for all runs, and find that 
they both change by less than about 10\% 
between $t=20$ and the end of the runs at $t=80$, giving the final values in Table \ref{T1}.
One can see that there is a linear relation $f = f_0 + f_1/Re_{\rm M}$ between the two, 
and a least squares fit gives $f_0 = 0.734 \pm 0.002$, $f_1 = -26.6 \pm 0.4$.

\begin{figure}[t!]
\centering{
\resizebox{8.5cm}{!}{\includegraphics{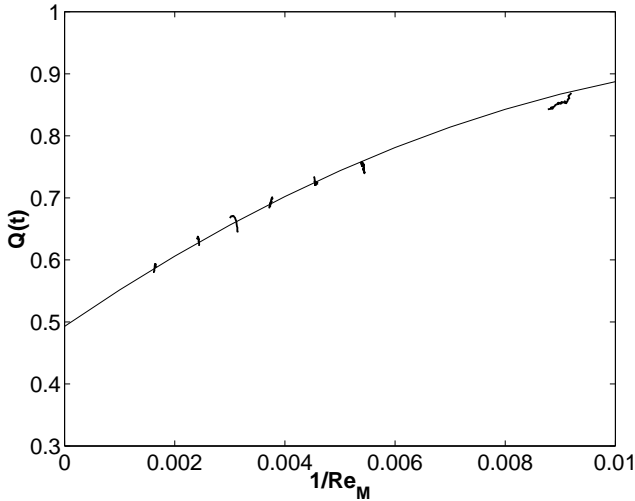}}
}
\caption{Energy decay exponent $Q$, defined in Eq.\ (\ref{RR}), plotted against 
the inverse magnetic Reynolds number $Re^{-1}_{\rm M}$,
calculated using the helicity scale $\xi_{\rm H}$, for the last half of the runs listed in Table 
\ref{T1}. The fit is a quadratic least squares fit using the final values for each run.
}
\label{f:Q_invRe_fit}
\end{figure}

The linear relation between $f$ and $Re^{-1}_{\rm M}$ implies that there should be 
a linear relation between $s$ and $Re^{-1}_{\rm M}$, and hence a quadratic relation between $Q(t)$, the energy decay exponent defined in Eq.\ (\ref{RR}), and $Re^{-1}_{\rm M}$.
Fig.~\ref{f:Q_invRe_fit}, showing $Q$ and $Re^{-1}_{\rm M}$  for $40 < t <80$, 
confirms that this is indeed the case, with asymptote at large $Re_{\rm M}$ consistent with 
$Q=1/2$. Furthermore, it also shows that both $Q$ and $Re^{-1}_{\rm M}$ change little over the last half of the runs.

Let us now turn to the decay of magnetic energy
which we use to show that the characteristic length scale of the field $\xi$ must
scale as $t^{1/2}$, assuming the scaling form of the magnetic energy power
spectrum
\ben
E_{M}(k) = \xi^{-q}g_{M}(k\xi)
\label{sca_law}
\een
found in Christensson et al.\ (2001). 
 From the MHD equations one can show 
\ben
  \dot E_{M} 
 = -\int d^{3}x\,\bu\cdot(\bJ\times\bB) - \eta\int d^{3}x\,\bJ^{2}. 
\label{energy_loss} 
\een
We assume that the dissipation term, if not dominant, always contributes a 
constant fraction to the energy loss.
If
$E_{\rm M}(k) \sim k^{-z}$ at high $k$, with $z<3$, then
the integral is dominated by the high-$k$ cut-off $\xi_{\rm diff}^{-1}$.
We do indeed
see a power spectrum with $ 2 < z < 3$ (Christensson et al.\ 2001), 
which means that there is a
well-defined helicity decay law, and that the energy decay is dominated by the
smallest scales.
Hence the dissipative loss $\ep_{M} =
-\dot E_{M}$ is then
\ben
\ep_{M} \sim \xi^{-q-z}\xi_{\rm diff}^{-3+z}.
\een
However, on integrating the energy power spectrum and then differentiating with
respect to time,
\ben
\ep_{M} \sim \dot\xi\xi^{-q-2} \sim t^{-1}\xi^{-q-1}.
\een
If we use the fact that $\xi_{\rm diff} \sim t^{1/2}$, we find on equating
the $\ep_{M}$ found in the two different ways, 
$
\xi^{-q-z} \sim t^{-(1+z)/2}\xi^{-q-1},
$
or
\ben
\label{e:LenScaGroLaw}
\xi \sim t^{1/2}.
\een
In deriving the power law for the length scale of the magnetic field, it
is not necessary that the dissipative terms account for all the energy loss,
just that it scales the same way as the energy loss and hence contributes 
a fixed fraction to the total. 

From the scaling form of the energy power spectrum (\ref{sca_law}), the energy decay 
law (\ref{energy_dec}) and the length scale growth law (\ref{e:LenScaGroLaw}) one
can easily show that a causal power spectrum with $n=2$ behaves as 
\ben
E_M(k,t) \sim k^4 t^{2-2s}
\een
at low $k$.  This growth at large scales should be taken into account 
when calculating the strength of primordially-produced fields, and will 
weaken the strong bounds derived by 
Durrer et al.\ (1998, 2003), Caprini \& Durrer (2001), Caprini et al.\ (2003).

\section{Comparison with 2D fluid turbulence}
\label{s:TwoDTurb}
There is an important similarity between 3D MHD and 2D fluid turbulence:
both exhibit inverse cascade behavior associated with the existence of
ideal invariants 
(Biskamp 1993)
In 2D fluid turbulence the energy shows an inverse cascade 
(Frisch \& Sulem 1984), 
which is similar to the inverse cascade of magnetic helicity in 3D MHD turbulence.
There are also some corresponding similarities between the decay laws in
decaying MHD turbulence and decaying 2D fluid turbulence.
Chasnov (1997) 
studied 2D velocity fields initialized with a Gaussian distribution, 
whose energy spectrum $E(k)$ is a power law at low $k$ and has an exponential cut-off at 
high $k$, much as we do.  

From the Navier-Stokes equation for an incompressible fluid one can easily show
\ben
\frac{\pa\vev{\bu^2}}{\pa t}  = -2 \nu \vev{\bomega^2},\quad
\frac{\pa\vev{\bomega^2}}{\pa t}  = -2 \nu \vev{\nabla\bomega^2},
\een
where $\bomega = \bna \times \bu$ is the vorticity (which is a scalar in two dimensions,
i.e.\ $\bomega=\omega\hat{\mathbf{z}}$).
We define two length scales from the enstrophy, $\vev{\bomega^2}/2$, and the palinstrophy, $\vev{\nabla\bomega^2}/2$,
\ben
l^2 = \vev{\bu^2}/\vev{\bomega^2}, \qquad l^2_{\rm p}=\vev{\bomega^2}/\vev{\nabla\bomega^2},
\een
in addition to the diffusion length scale $l_{\rm diff} = \sqrt{\nu t}$. Postulating power-law 
decays for the energy and enstrophy, 
\ben
\vev{\bu^2} \sim t^{n}, \qquad \vev{\bomega^2} \sim t^{m},
\een
we find that $m = n-1$, and that 
\ben
n = - 2 \nu t \vev{\bomega^2}/\vev{\bu^2} = - 2 (l_{\rm diff}/l)^2.
\een
Hence, a power-law decay again requires that $l \sim t^{1/2}$, 
and in the limit that the length scale of the flow becomes large compared with the diffusion 
scale, the energy is conserved.  This is analogous to the conservation of magnetic 
helicity in 3D MHD.

However, the Reynolds number, approximately constant in our simulations, 
behaves rather differently in 2D fluid turbulence. 
Defining $Re = u l/\nu$, with $u = \vev{\bu^2}^{1/2}$, we find $Re \sim t^{(1+2)/2}$, 
at least for large Reynolds numbers.  For small Reynolds numbers we expect 
$Re$ to decay to zero through dissipation, pointing to the existence of a critical 
Reynolds number $Re_c$, at which the system can remain for long periods, if 
initialized close enough to the correct value (Chasnov 1997). 

The energy spectrum of decaying turbulence is also found to be self-similar in 2D.
Similar arguments to those presented in Section \ref{helicity_conservation}
show that self-similarity again implies $l\sim t^{1/2}$.
These arguments also show that if the palinstrophy length scale $l_{\rm p} $
is of the same order of magnitude as 
the enstrophy length scale $l$, and that enstrophy is being 
dissipated at the diffusion scale $l_{\rm diff}$, 
the high-$k$ exponent of a power-law energy spectrum must be between $-4$ and $-5$.
Intriguingly, the results of Chasnov (1997)
seem to indicate a power law somewhat steeper 
than the $k^{-3}$ behavior expected in stationary turbulence 
{Kraichnan \& Nagarajan 1967, Batchelor 1969}.

\section{Discussion and conclusions}
\label{s:disc}

We have studied the evolution of decaying 3D MHD turbulence involving 
maximally helical magnetic fields. For finite magnetic diffusivity there emerges  
an important quantity $s = (\xi_{\rm diff}/\xi_{\rm H})^2$, where $\xi_{\rm H}$ 
is the helicity scale defined in Eq.\ (\ref{length_H}), and $\xi_{\rm diff}$ is the diffusion 
scale.  We find $\xi_{\rm H} \simeq vt$, where $v$ is the RMS velocity, and hence that 
$s \propto Re^{-1}_{\rm M}$, the magnetic Reynolds number evaluated using the 
helicity scale.  
The magnetic field coherence length (which can be equally well expressed 
as the integral, helicity or relative helicity scales) goes as
$\xi \sim t^{1/2}$, magnetic helicity $H_{\rm M} \sim t^{-2s}$ and
magnetic energy $E_{\rm M} \sim t^{-1/2 - 2s}$.  
A corollary is that $Re_{\rm M}$ is constant once the system has 
reached self-similarity.
Furthermore, we can extrapolate to the limit of
very large magnetic Reynolds numbers, useful for example in the early
Universe, to find $H$ constant and $E_{\rm M} \sim t^{-1/2}$.

Our model for the scaling laws should be compared with that of
Biskamp \& M\"uller (1999). 
The first difference is that they assumed that 
the non-linear term in the evolution equation for the magnetic field was the 
dominant source of energy loss for the magnetic field, and that the magnetic 
field was asymptotically the dominant contributor to the total energy $E$, 
expressed as 
$\Gamma \equiv E_{\rm V}/E_{\rm M} \ll 1$.  Then we can write
\ben
\dot E \sim \Gamma^\half E^{3/2}/\xi,
\een
where $\xi$ is a length scale of the magnetic field.  They then found the 
phenomenological relation $\Gamma \simeq E/H$, which, when coupled with 
$E \simeq H/\xi$ and the conservation of $H$, gives $E \sim t^{-1/2}$.

We emphasize that this model is not 
inconsistent with ours. 
We infer $E_{\rm V} \sim t^{-1}$ from the relation $\xi_{\rm H} \simeq vt$, 
and hence that $\Gamma \sim t^{-1/2+2s} \sim (E/H)t^{2s}$.  The difference 
between Biskamp \& M\"uller's assumed relation $\Ga \sim E/H$ and ours 
is small at large magnetic Reynolds numbers where $s\to 0$.
Furthermore, both approaches need to assume only that the 
non-linear and dissipative terms in Eq.\ (\ref{energy_loss}) are not 
sub-dominant (rather than dominant) and it turns out that both scale 
with time in the same way, as $E_{\rm M}/t$.  In our simulations dissipation typically 
accounted for about 60\% of the energy loss in the period  
$t\simeq 40$ to $t \simeq 80$,
which means that the field is not force-free.

We believe that our model has certain advantages, in
that the assumptions going into it give
more physical insights than the phenomenological 
(and dimensionally incomplete)
relation $\Ga \sim E/H$.  
Our assumptions are
that the magnetic power spectrum exhibits a self-similar form (\ref{sca_law}),
with power-law behavior $k^{-z}$ at high $k$, that resistive dissipation occurs
predominantly at the diffusion scale $\xi_{\rm diff}$, that there is a separate
helicity scale $\xi_{\rm H}$, from which it follows that $2<z<3$.  We also
assume that the eddy turn-over time $\xi/v$ is $t$, and that 
the relative helicity is asymptotically constant, i.e.\ that the energy 
decays as fast as possible as is consistent with helicity conservation.  
Given these assumptions, it
follows that the dissipative and non-linear terms in the magnetic energy loss
equation (\ref{energy_loss}) scale the same way, and we can infer that $\xi \sim
t^{1/2}$.  From this we derive a scaling law for $H$, 
finding as expected that $H$
is conserved in the limit of large magnetic Reynolds number, and with the assumption of 
constant relative helicity, we obtain that $E \sim t^{-1/2}$ in the same limit.

In summary, for both 3D helical MHD and 2D fluids, 
the key to understanding the 
free decay of turbulence is self-similarity, coupled to a separation between the scale 
of the flow and the diffusion scale.  

\textit{Note added.\ } Since the first draft of this work appeared
as astro-ph/0209119,v1, there 
have been several developments. A new review of magnetic 
fields in cosmology has appeared (Giovannini 2004). 
Using different scaling arguments, Campanelli (2004) 
has rederived our scaling laws for decaying helical 
turbulence, and extended the analysis to the non-helical case, where the ideal 
limit has $E_{\rm M} \propto t^{-1}$, $E_{\rm V} \propto t^{-1}$, and $\xi \propto t^{1/2}$. 
Extensive 3D numerical simulations, including the interesting case of Prandtl number 
larger than unity, have been carried out by Banerjee \& Jedamzik (2003, 2004).
The authors did not seem to be aware of our work or of
Campanelli (2004) 
and did not attempt to analyse  the 
decay of the helicity, kinetic or magnetic energies in its light.  However, examination 
of their Figs.\ 1 and 7 in the later paper show decay laws 
close to $t^{-0.5}$ and $t^{-1}$ for the helical 
and non-helical cases respectively (with Prandtl number unity).
They do not quote Reynolds numbers for the simulations so an exact comparison 
cannot be made.
There has also been further work on effects of primordial fields on the CMB
(Kosowsky et al.\ 2005, Kahniashvili \& Kahniashvili 2005)
and on the evolution of magnetic fields in the 
post-recombination era (Sethi \& Subramanian 2004). 

\acknowledgements
This work was conducted on the Cray T3E and SGI Origin platforms using COSMOS 
Consortium facilities, funded by HEFCE, PPARC and SGI. We also acknowledge
computing support from the Sussex High Performance Computing Initiative. 
MH thanks for NORDITA for hospitality, and MC the Astronomy Centre at the University 
of Sussex.

\newcommand{\ybook}[3]{ #1, {\it #2} (#3)}
\newcommand{\yprl}[3]{ #1, {PhRvL,} {#2}, #3}
\newcommand{\yprd}[3]{ #1, {PhRvD,} {#2}, #3}
\newcommand{\ypre}[3]{ #1, {PhRvE,} {#2}, #3}
\newcommand{\yprt}[3]{ #1, {PhR,} {#2}, #3}
\newcommand{\yjour}[4]{ #1, {#2}, {#3}, #4}
\newcommand{\yapj}[3]{ #1, {ApJ,} {#2}, #3}
\newcommand{\yan}[3]{ #1, {AN,} {#2}, #3}
\newcommand{\yjfm}[3]{ #1, {JFM,} {#2}, #3}
\newcommand{\yplb}[3]{ #1, {PhLB,} {#2}, #3}
\newcommand{\ypf}[3]{ #1, {PhLB,} {#2}, #3}
\newcommand{\ymn}[3]{ #1, {MNRAS,} {#2}, #3}
    
\end{document}